\documentclass[12pt,preprint]{aastex}
\usepackage{rotating}
\usepackage{multirow}
\usepackage{color}
\usepackage{CJK}

\slugcomment{}
\begin{document}

\title{Evidence of Different Formation Mechanisms for Hot versus Warm Super-Earths}
\begin{CJK*}{UTF8}{gkai}

\author{Wei Zhu (祝伟) \\
%\author{Wei Zhu \\
\normalsize{Department of Astronomy, Ohio State University, 140 W. 18th Ave., Columbus, OH  43210, USA}
}

\email{weizhu@astronomy.ohio-state.edu}

%\altaffiltext{}{Department of Astronomy, The Ohio State University, 140 W. 18th Ave., Columbus, OH 43210, USA}

\begin{abstract}
    Using the Kepler planet sample from Buchhave et al. and the statistical method clarified by Schlaufman, I show that the shorter-period super-Earths have a different dependence on the host star metallicity from the longer-period super-Earths, with the transition period being in the period range from 70 to 100 days. The hosts of shorter-period super-Earths are on average more metal-rich than those of longer-period super-Earths. The existence of such a transition period cannot be explained by any single theory of super-Earth formation, suggesting that super-Earths have formed via at least two mechanisms.
\end{abstract}


One of Kepler's main discoveries is the existence of abundant planets with sizes from Earth ($R_\oplus$) to Neptune ($4R_\oplus$) \citep{Borucki:2011a,Borucki:2011b,Batalha:2013,Burke:2014}. These so-called ``super-Earths'' pose challenges to the standard core accretion model \citep{IdaLin:2004}. Various models have been proposed to explain the formation process of such planets \citep{IdaLin:2010,HansenMurray:2012,ChiangLaughlin:2013}. Although the origin of giant planets (planetary radius $R_{\rm p}>4R_\oplus$) has been reasonably well understood, the formation mechanism for super-Earths is still under debate. A class of models suggest that these planets are formed \textit{in situ}, with no significant post-assembly migration \citep{HansenMurray:2012,ChiangLaughlin:2013}. Other models suggest that they may have a similar formation mechanism as giant planets, namely forming beyond the snow line where the solid material is rich and then migrating in via various interactions \citep{TerquemPapaloizou:2007,IdaLin:2010}. These two classes of models diverge in terms of the dependence on the richness of rocky material: the \textit{in situ} formation mechanism generically requires an unusually large amount of solids in the inner region of the protoplanetary disk, while the alternate theory does not have this restriction. Therefore, a natural way of distinguishing these two theories is to look at the planet-metallicity relation of super-Earths as a function of orbital period.

\begin{figure}
\centering
\plotone{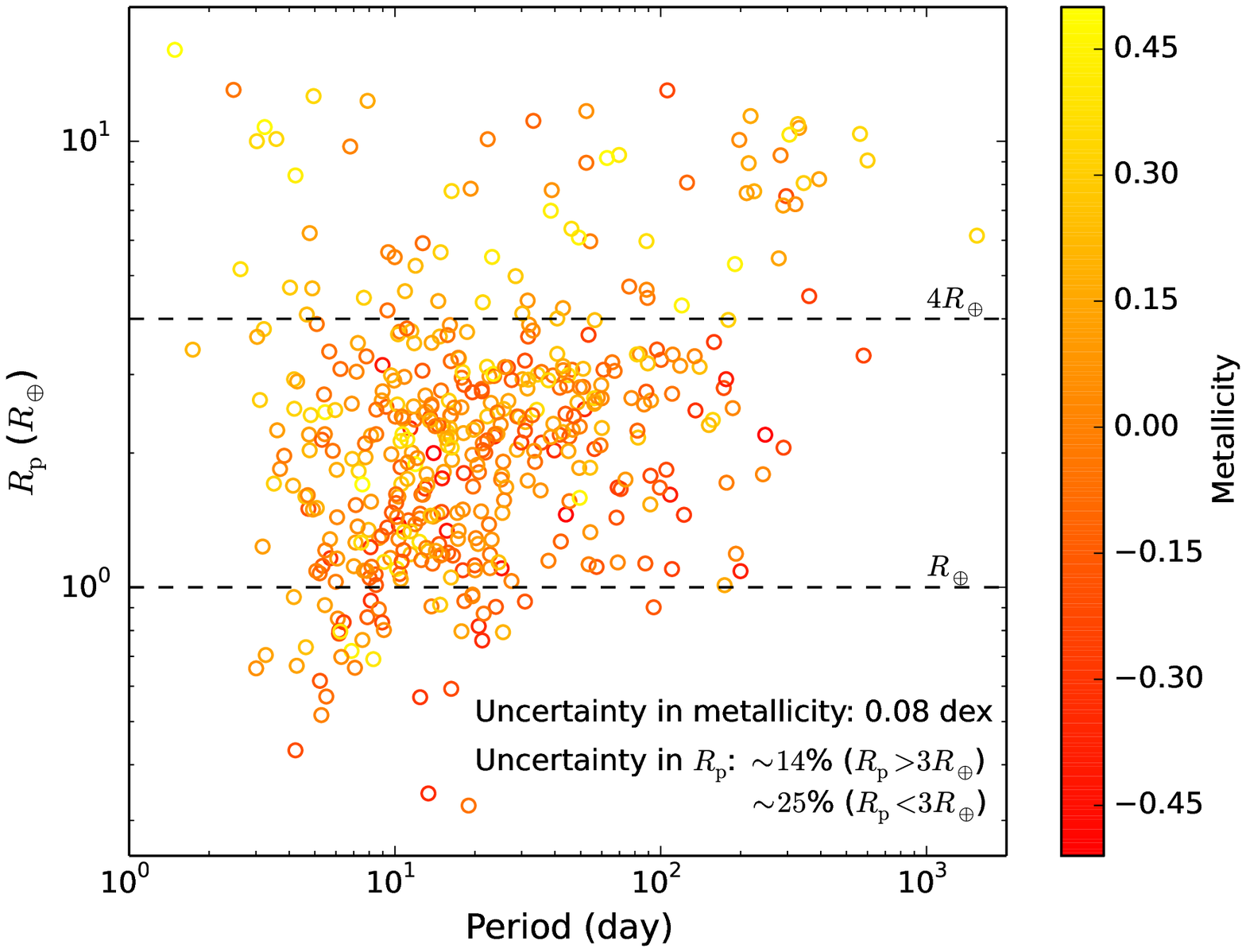}  
\caption{Planet radius $R_{\rm p}$ vs. orbital period of the final planet sample; data points are color-coded by the host star metallicities.}
\label{fig:Rp-period}
\end{figure}

The planet (candidate) sample used in the present work is taken from a subgroup of the Kepler sample, which has been used to identify the transition radius from gaseous to rocky planets \citep{Buchhave:2014,Schlaufman:2015}. The overall sample consists of 600 exoplanet candidates orbiting a total of 405 Solar-like stars (F, G and K spectral types). The stellar parameters were determined based on high-resolution spectra gathered by the Kepler Follow-up Program.
\footnote{See the Methods section of the original paper\citep{Buchhave:2014} for the details of the observations and data reduction.}
Following the two previous studies \citep{Buchhave:2014,Schlaufman:2015} based on the same dataset, I also exclude small ($R_{\rm p}<3R_\oplus$) planets that are highly irradiated (stellar flux $F>5\times10^5{\rm\ J\ s^{-1}\ m^{-2}}$) from the sample, because these planets could have undergone significant evaporation of their atmospheres and thus no longer follow the primordial planet-metallicity relation \citep{OwenWu:2013}. This shrinks the sample by $23\%$, yielding 463 planets orbiting 324 stars. To account for the uncertainties in transit depths, which are significant for small planets \citep{Schlaufman:2015}, the error bars on the planetary radii of small planets ($R_{\rm p}<3R_\oplus$) are inflated by a factor of two with respect to the originally reported radii. The orbital period of each planet is taken from the NASA Exoplanet Archive
\footnote{\url{http://exoplanetarchive.ipac.caltech.edu/}.},
which come from the Kepler pipeline \citep{Borucki:2011a,Borucki:2011b,Batalha:2013,Burke:2014}. The sizes and orbital periods of the planets in the final sample are shown in Figure~\ref{fig:Rp-period} as a scatter plot, color coded by the host star metallicity. Typical uncertainties are $\sim 14\%$ and $\sim 25\%$ on the planetary radii of large and small planets, respectively, and 0.08 dex on stellar metallicities; planet orbital periods are very well constrained (fractional error $<0.001\%$), so this uncertainty is not considered hereafter.

Two populations of planets are chosen for further analysis: giant planets ($R_{\rm p}>4 R_\oplus$) and super-Earths ($R_\oplus<R_{\rm p}<4R_\oplus$). Planets smaller than Earth are not included in the smaller planet sample because studies on planet occurrence rates suggest that these sub-Earth sized planets may have formed through a different channel \citep{Petigura:2013}; they are not considered as a separate population either, because only a few long-period such planets are detected in the current sample.

\begin{figure}
\centering
\epsscale{1.1}
\plottwo{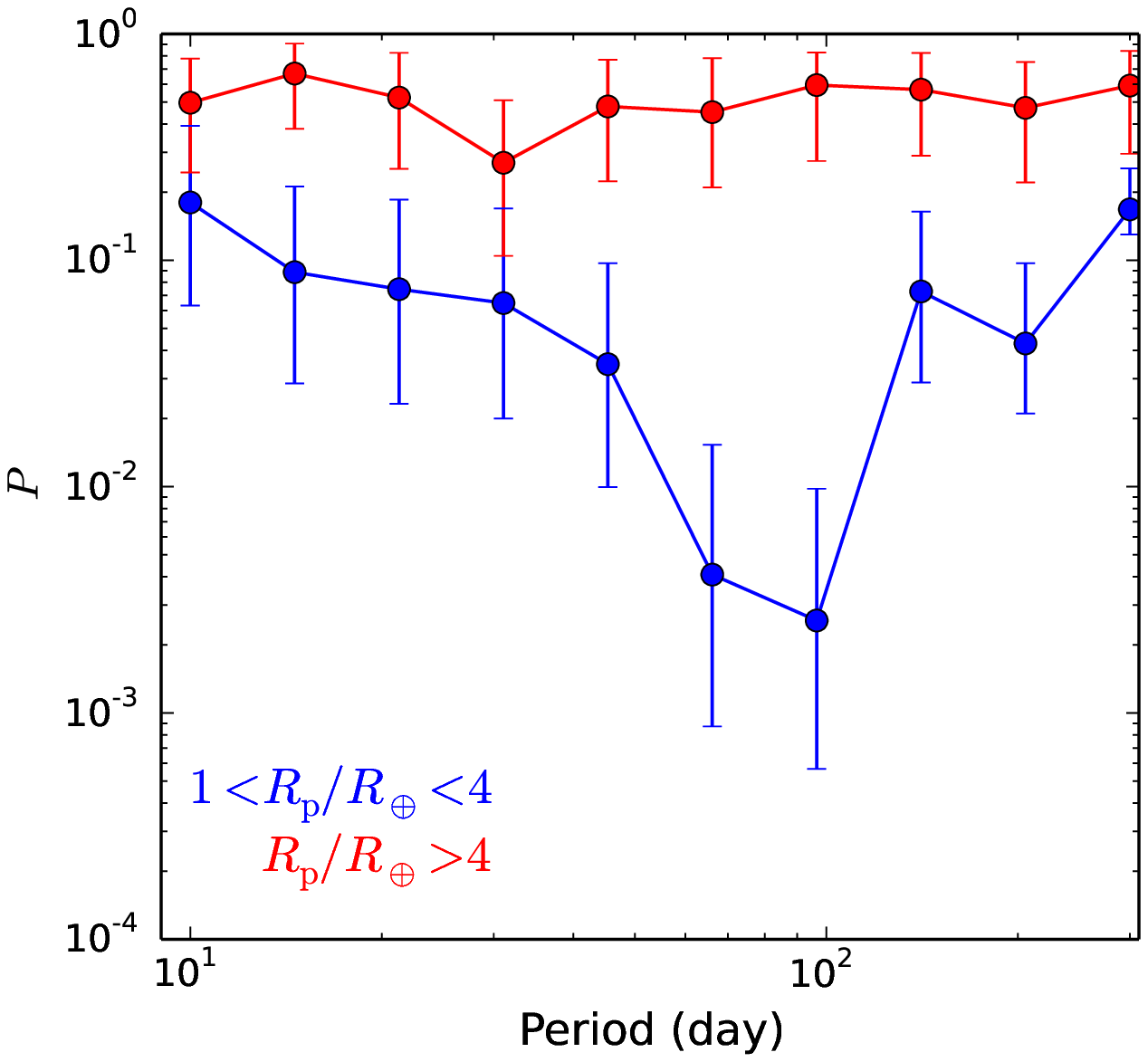}{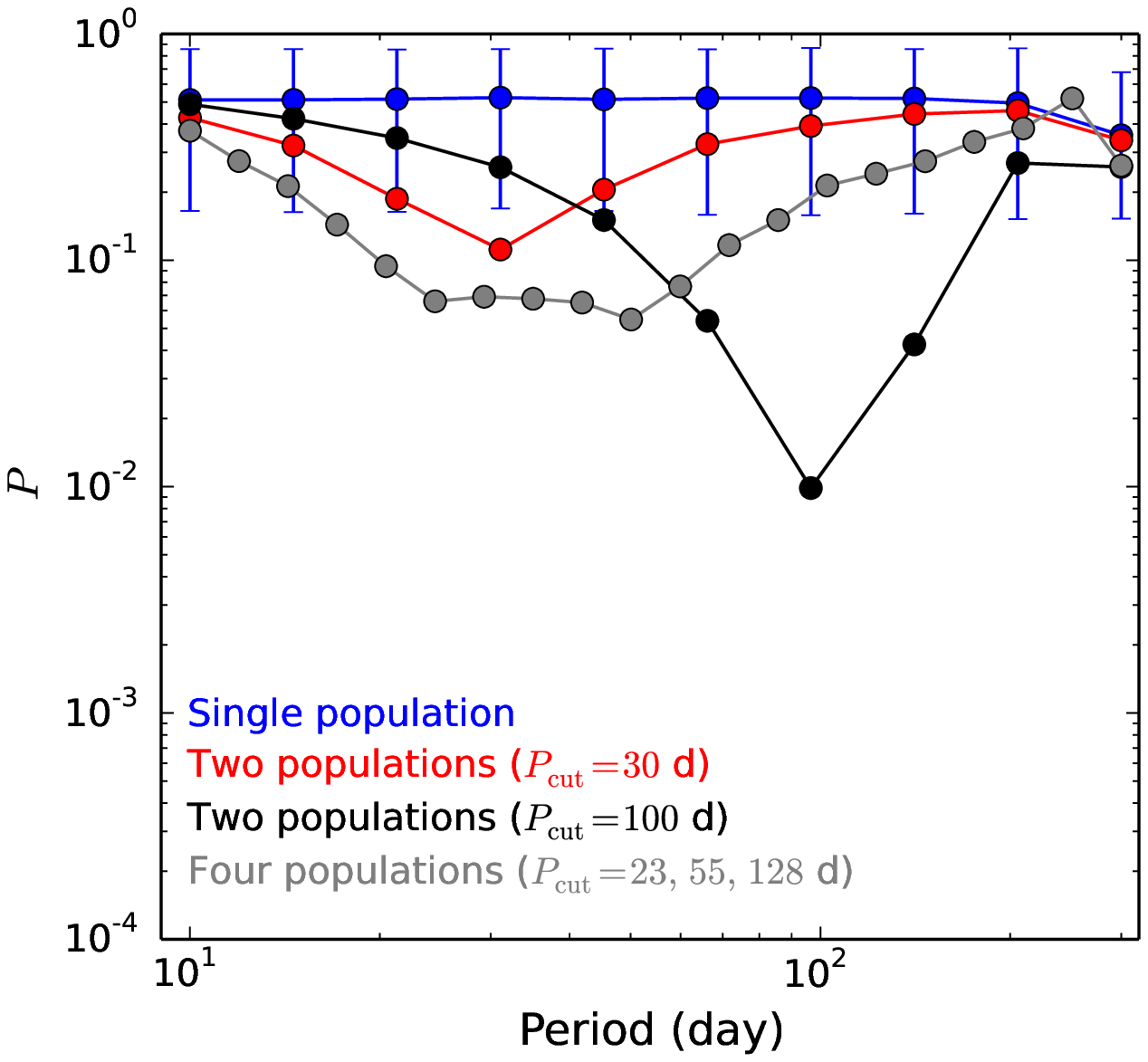}
\caption{\textit{Left panel}: the $p$ values of the two-sample Kolmogorov-Smirnov test at various orbital periods, for giant planets (red) and super-Earths (blue) in the \citet{Buchhave:2014} planet sample. \textit{Right panel}: results of the signal injection and recovery test; the blue curve is the result assuming a single super-Earth population. Red and black curves are the recovery tests of two populations, with transition period at 30 d (red) and 100 d (black); the gray curve shows a four-population test, which represents smooth metallicity-period relation; for clarity, only the blue curve shows error bars.}
%The $p$ values of the two-sample Kolmogorov-Smirnov test at various orbital periods, for real data (\textit{left panel}) and fake data (\textit{right panel}). giant planets (red) and super-Earths (blue). The flat distribution for giant planets indicates that there is only one giant planet population within the probed period range, while the local minimum appearing in the super-Earths case suggests that shorter-period ($P \lesssim 100$ d) super-Earths have different metallicity dependence from their longer-period counterparts.}
\label{fig:pval-period}
\end{figure}

\begin{figure}
\centering
\epsscale{0.8}
\plotone{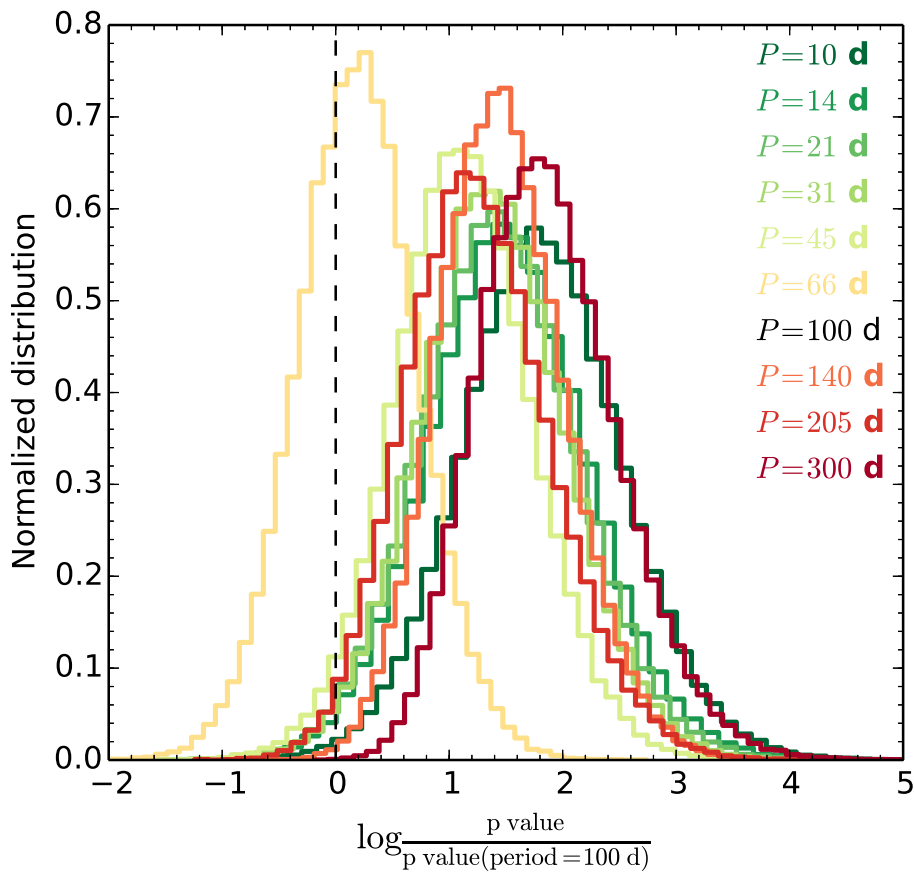}
\caption{The differences between $p$ values at other orbital periods and at 100 day from the $10^5$ realizations. In large majorities of the $10^5$ realizations, the $p$ value at $P=100$ d is always significantly smaller than $p$ values at other periods except $P=66$ d, and the $p$ value at $P=66$ d is on average equally small as that at $P=100$ d. These results indicate that the local minimum is indeed significant, and lies near the interval $P=66$ to $100$ d.}
\label{fig:pval-diff}
\end{figure}

\begin{figure}
\centering
\plotone{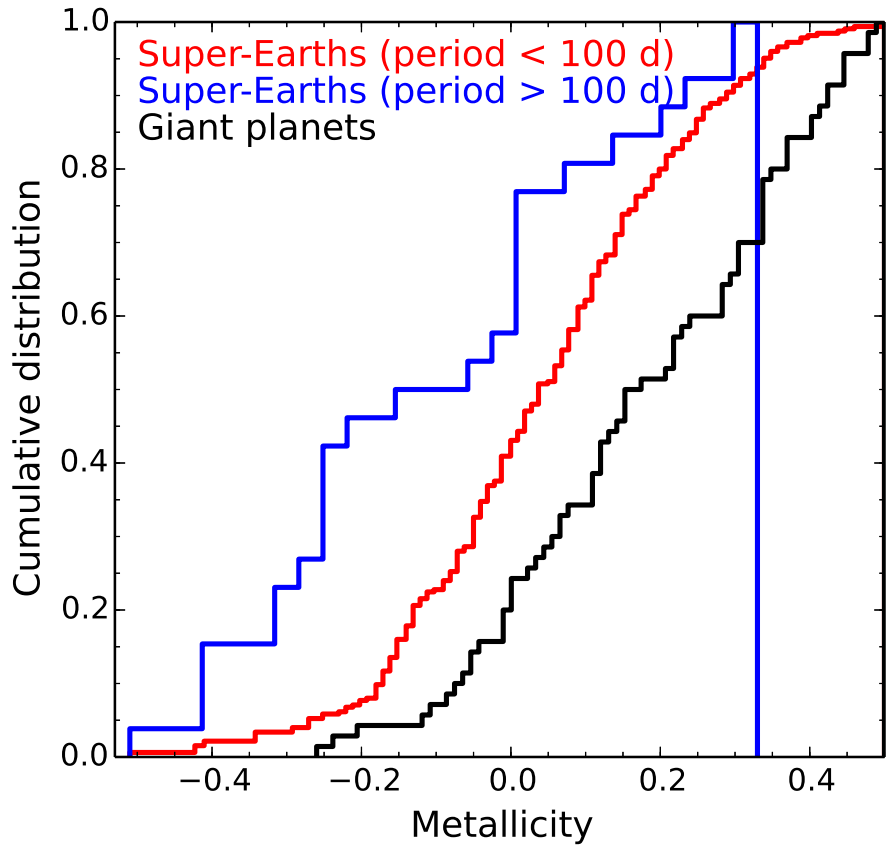}
\caption{The cumulative distributions of host star metallicity for giant planets (black) and super-Earths with orbital period shorter than (red) and longer than (blue) 100 days. The average metallicities for giant planets and the two super-Earth populations are $0.17\pm0.02$, $0.04\pm 0.01$ ($P<100$ d), and $-0.09\pm 0.04$ ($P>100$ d).}
\label{fig:cdf-z}
\end{figure}

For each planet population, the two-sample Kolmogorov-Smirnov (2KS) test is then used to identify whether there is a transition period. A similar method has been used to identify the transition radius from gaseous to rocky planets \citep{Buchhave:2014,Schlaufman:2015}. Given an orbital period, each planet population is then divided into two groups consisting of shorter-period and longer-period planets, and then the host star metallicity distributions of the two groups are compared, and the difference is quantified by the 2KS test $p$ value. This procedure is repeated for 10 period values from 10 to 300 days equally separated in log scale. To account for the uncertainties in planetary radii and stellar metallicities, I take $10^5$ realizations of the dataset, in each of which the planet radii and host star metallicities are randomly drawn from the Gaussian distributions defined by the measured values and given error bars. The $p$ values at each orbital period are computed for each realization, and the final reported $p$ value for this orbital period is taken as the median, with uncertainty defined by the $16$ to $84$ percentile values of the cumulative distribution of $p$ values from $10^5$ realizations. As shown in the left panel of Figure~\ref{fig:pval-period}, the $p$ values at various orbital periods are essentially the same for giant planets, suggesting no distinct formation mechanisms for giant planets within the period range that is probed by the current sample. However, a local minimum of the $p$ value distribution is found at $\sim 100$ days for super-Earths. Given the uncertainties in the $p$ values, this local minimum does not appear at first sight prominent. However, these error bars are highly correlated, as demonstrated by Figure~\ref{fig:pval-diff}, which shows the distribution of differences between $p$ values at other periods and at $P=100$ d. This shows that the minimum near $P=100$ d is indeed significant. Figure~\ref{fig:pval-diff} also shows that the $p$ value at $P=66$ day is on average equally small as that at $P=100$ day, which suggests that the transition occurs somewhere in this range. I choose 100 day for illustration, and show the cumulative distributions of host star metallicities for the two groups of super-Earths separated by this orbital period in Figure~\ref{fig:cdf-z}. For short-period ($P<100$ d) and long-period ($P>100$ d) super-Earths, the mean metallicities of host stars are $0.04$ and $-0.09$, respectively, with uncertainties of $0.01$ and $0.04$ in these mean values. The cumulative distribution of host star metallicities for all giant planets is also shown in Figure~\ref{fig:cdf-z}, with mean metallicity $0.17$ and an uncertainty $0.02$, consistent with previous studies \citep{FischerValenti:2005,Buchhave:2012}.

I conduct the following tests, first to verify the observed local minimum in the $p$ value distribution is physically real rather than the result of a random sampling effect, and then to investigate whether it is caused by two distinct super-Earth populations or a smooth transition of the metallicity-period correlation.
I first assume a single population of super-Earths, with host star metallicity of each planet randomly drawn from a Gaussian distribution [$\mathcal{N}(0.03,0.19^2)$, with $0.03$ the mean and $0.19$ the standard deviation]. Then I repeat the Monte Carlo process as described above to produce the $p$ value \textit{vs.} period plot, which is shown as blue points with error bars in the right panel of Figure~\ref{fig:pval-period}. I then assume two different populations of super-Earths, first with a transition period $P=30$ d and then with $P=100$ d, the results of which are also shown in the right panel of Figure~\ref{fig:pval-period}. In both cases the injected transition periods can be recovered reasonably well. The right panel of Figure~\ref{fig:pval-period} also shows the result from the case with four different populations, with the three transition periods (23 d, 55 d, and 128 d) equally separated in log scale. Given the size of the planet sample, the four-population case represents a smooth transition on the metallicity-period relation, in contrast with the rapid transition for the two-population case. Compared with the result based on real data (the left panel of Figure~\ref{fig:pval-period}), the two-population model is favored over the smooth transition one.

\begin{figure}
\centering
\plotone{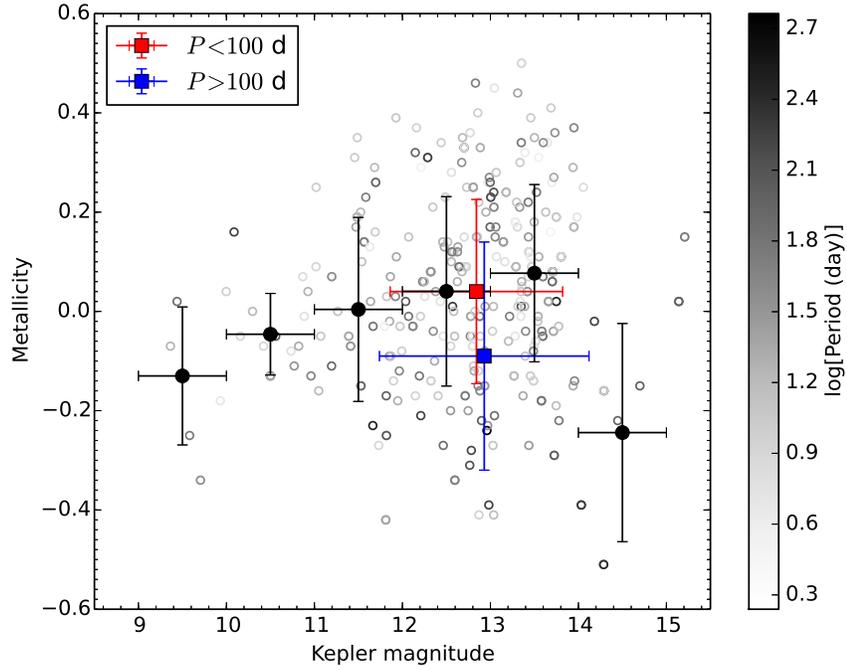}
\caption{The selection bias in the sample \textit{vs.} the observed trend for super-Earths. Points without error bars are the positions of the super-Earth host stars in the metallicity \textit{vs.} Kepler magnitude (brighter to fainter), color coded by the planet orbital period. This sample is grouped into six magnitude bins, with the vertical error bar representing the standard deviation of the stellar metalliciteps within each bin. The centroids for two super-Earth populations are marked with red (super-Earths with $P<100$ d) and blue (super-Earths with $P>100$ d).}
\label{fig:bias}
\end{figure}

I address the influence of the systematic and selection effects before proceeding to discuss the theoretical implications of the results. Since Kepler planets are selected partly based on the signal-to-noise ratio, longer-period planets can be more easily found around brighter stars, which might bias on metallicities since stars with different metallicities enter the sample with systematically different observed fluxes. However, this selection effect turns out to be in the opposite direction as it would have to be in order to explain the results. In Figure~\ref{fig:bias} I show the scatter plot of the metallicity \textit{vs.} Kepler magnitude of all super-Earth hosts, with colors coded by the orbital periods of surrounding planets. All targets are grouped into six magnitude bins, which show a tendency of gradually increasing in metallicity as stars become fainter. The two super-Earth populations, however, show an opposite effect: longer-period super-Earths tend to have hosts that are slightly fainter but more metal-poor than the shorter-period ones. Therefore, the inclusion of the selection effect can only make the two super-Earth populations more distinct.

%but this is primarily a geometric bias. In a color-based survey such as Kepler, more metal-poor stars turn to be brighter, so the transit planet sample may turn to have more metal-poor stars, while on the other hand, the requirement of high-resolution spectra would presumably result in more metal-rich stars in the sample. Although it is not clear how the final sample is compared to a controlled sample of all field stars, the bias on stellar metallicity is not likely to correlate with the orbital period of planet. Therefore, I suggest that the detected transition period for super-Earths is probably originated from the physical processes that these planets may have undergone.

The results suggest that close-in super-Earths are more often found around more metal-rich stars, while widely separated super-Earths are more often around more metal-poor stars. What does it imply for the formation mechanisms of super-Earths? The \textit{in situ} formation scenario seems to be favored over the migration scenario, 
since more metal-rich stars would presumably have more solid material at shorter distance in the protoplanetary disk and thus it would be easier to form super-Earths where they are now. However, the \textit{in situ} formation scenario fails to solely explain the observed tendency for two reasons. First, if the \textit{in situ} formation holds for all super-Earths, one would expect to see a gradually weakening dependence on stellar metallicity as the planet orbital period increases, rather than the very distinct transition period as is observed here. Another reason that also makes the \textit{in situ} super-Earth formation very difficult is that the overall solid material required to form such planets is so large that it almost exceeds what is available from the protoplanetary disk could have at such distances \citep{HansenMurray:2012,ChiangLaughlin:2013}. For example, the minimum-mass solar nebula\citep{Hayashi:1981} can only provide $\sim 3.3 M_\oplus$ solids interior to 1 AU. However, there are other arguments that may make the second reason unimportant, such as the efficiency in converting nebular material into planets being very high or most of these super-Earths consisting of a large amount of gas envelope.
There are also theories that explain the super-Earths as the failed cores of giant planets. However, such theories also have the problem of explaining the position of the observed transition period. 

Although a transition period is not detected in the current sample for giant planets, it may be that such a transition period, if it exists, is significantly beyond 300 days. Since giant planets are believed to have formed near the ice line ($\sim3$ AU for Solar-like stars) where plenty of solid material is accessible for building up the massive core \citep{Lin:1996}, one can speculate that the possible transition period may be of order thousands of days. Therefore, the detection of the transition period of giant planets is currently inaccessible. However, since the number of planets detected via radial velocity technique is growing and extending to longer periods, this putative transition point may become accessible in the near future. In addition, microlensing provides another approach to answer this question. Being sensitive to planets near and outside the ice line, microlensing is probing a very different planet population from other techniques \citep{MaoPaczynski:1991,GouldLoeb:1992,Gould:2010,Sumi:2011}. If the microlens parallax effect can be routinely measured for a large fraction of all planetary events, the planets detected via microlensing will have well constrained mass and distance measurements \citep{Udalski:2015,CalchiNovati:2014}. When combined with the Galactic distribution of stellar metallicities, it in principle could also yield a planet-metallicity relation for very long-period giant planets.

\acknowledgments
I thank Andrew Gould for insightful discussions, and Andrew Gould and Scott Gaudi for careful reading of the manuscript. This work is supported by JPL (Spitzer) grant 1500811. Some of the data used in this work were collected by the Kepler mission. Funding for the Kepler mission is provided by the NASA Science Mission Directorate.

%\bibliographystyle{apj}
%\bibliography{references}

\clearpage

%\begin{figure}
%\centering
%\plotone{Zhu_f5.eps}
%\caption{Results of the signal injection and recovery test, shown on the same scale as Figure~\ref{fig:pval-period}. The blue curve is the result assuming a single super-Earth population. Red and black curves are the recovery tests of two populations, with transition period at 30 d (red) and 100 d (black); in both cases, the injected transition periods are recovered. The gray curve shows a four-population test, which represents smooth metallicity-period relation. For clarity, only the blue curve shows error bars.}
%\label{fig:recovery}
%\end{figure}

\end{CJK*}
\end{document}